\documentclass[%
superscriptaddress,
 amsmath,amssymb,
 aps,
reprint,
]{revtex4-2}

\usepackage{graphicx}
\usepackage{dcolumn}
\usepackage{bm}

\usepackage[utf8]{inputenc}
\usepackage[T1]{fontenc}
\usepackage{mathptmx}
\usepackage{etoolbox}

\usepackage{xcolor}

\makeatletter
\def\@email#1#2{%
 \endgroup
 \patchcmd{\titleblock@produce}
  {\frontmatter@RRAPformat}
  {\frontmatter@RRAPformat{\produce@RRAP{*#1\href{mailto:#2}{#2}}}\frontmatter@RRAPformat}
  {}{}
}%
\makeatother

\begin{document}

\preprint{AIP/123-QED}

\title{ A Gauge Field Theory of Coherent Matter Waves }
\author{Dana Z. Anderson}
\email{dana@infleqtion.com}
\affiliation{Infleqtion, 3030 Sterling Circle, Boulder, CO 80301, USA}
\altaffiliation[Also at ]{Department of Physics and JILA, University of Colorado, Boulder CO 80309-0440, USA}
\author{Katarzyna Krzyzanowska}
\affiliation{Materials Physics and Applications Division, Los Alamos National Laboratory, Los Alamos, NM 87545, USA }

\date{\today}

\begin{abstract}
A gauge field treatment of a current oscillating at frequency $\nu$ of interacting neutral atoms leads to a set of matter-wave duals to Maxwell's equations for the electromagnetic field.  In contrast to electromagnetics, the velocity of propagation has a lower limit rather than upper limit and the wave impedance of otherwise free space is negative real-valued rather than 377$\Omega$. Quantization of the field leads to the matteron, the gauge boson dual to the photon. Unlike the photon, the matteron is bound to an atom and carries negative rather than positive energy, causing the source of the current to undergo cooling. Eigenstates of the combined matter and gauge field annihilation operator define the coherent state of the matter-wave field, which exhibits classical coherence in the limit of large excitation.
\end{abstract}
 
\maketitle

\section{Introduction}
This work considers an oscillating current of interacting ultracold atoms through the lens of non-relativistic field theory. Over the past few decades,  relativistic quantum field theory has led to the unification of three of the four fundamental forces and now provides our modern picture of the physics of fundamental particles as the Standard Model \cite{cheng.1984}.  Non-relativistic systems, in particular the connection between Hamilton-Jacobi theory and quantum mechanics, have been studied using a field-theoretic approach since the early days of quantum mechanics \cite{whittaker_1941, BohmPhysRev1952}. In line with these early works and following the reasoning of Wheeler \cite{Wheeler.1995}, the problem of interacting identical neutral particles can be treated using classical field theory; utilized in conjunction with a Hamilton-Jacobi formalism the approach leads naturally to a wave description of particles and their interactions.  Indeed, it was a significant revelation of the last century that Maxwell's equations themselves follow from the treatment of electrons as excitations of a field taken together with certain symmetry considerations\cite{Soper.1976,Franklin.2017}. Our work draws on electromagnetism as a familiar application of classical field theory and offers a new as well as useful perspective on matter waves.

The wave characteristics of quantum-mechanical particles, either individual or ensembles, are commonly referred to as ``matter waves''; they are characterized by a de Broglie wavelength $\lambda=h/p$, where $h$ is Planck's constant and $p$ is the particle momentum. An excellent example of matter waves is the flux of atoms extracted from a Bose-Einstein condensate (BEC) \cite{MewesPRL1997, RyuNJoP2015} that was first achieved in 1995 in a diluted gas of rubidium and sodium atoms \cite{Anderson.1995bec, Davis.1995}. Early experiments demonstrated the quantum coherence of these systems \cite{Andrews.1997, BurtPRL1997, KetterlePRA1997}. The achievement of BEC prompted many-body physics to shift its focus towards ultracold gases, resulting in various models, analytical techniques, and numerical methods, such as the Gross-Pitaevskii equation (GPE), time-dependent Hartree-Fock-Bogoliubov (TDHFB), Bogoliubov-de Gennes equation, quantum kinetic theory, and stochastic methods, among others \cite{Pethick.2002,Stringari.2003,Griffin.2009,Blakie.2008cap}. 

In parallel with advances in atom cooling and manipulation techniques so have there been advances in atom interferometry \cite{Berman.1997, Kasevich.2016, Biedermann.2017, Lee.2018, Elliott.2018, Mazon.2019, Malinovsky.2019, Kumar.2020, Abend.2020, Thom.2021, Lachmann.2021, Li.2021} that take advantage of the properties of ultracold atoms, particularly their  coherence. Atom interferometry and its use of matter-waves has been very much inspired by its optical interferometry counterpart.  At the same time, the nature of BEC systems also leads one naturally to think in terms of atom currents flowing from one part of a system to another. Thus arose the field of atomtronics, the atom analog of electronics in which atom flux and chemical potential substitute for electric current and potential \cite{Seaman.2007guo,Pepino.2007,Pepino.2010,RyuNJoP2015,Amico.2017,AmicoAVSQ2021,Pepino.2021,Anderson.2021,Amico.2022}. 

As fundamentally non-thermal-equilibrium open quantum systems, even rather simple atomtronic circuits prove challenging for standard many-body methods. Yet we can look more deeply at the analogy between atomtronic currents versus electronic currents and matter waves versus electromagnetic waves. Maxwell's equations inform us that an oscillating electric current gives rise to a coherent electromagnetic wave.  We can ask, therefore, whether an oscillating atomtronic current gives rise to a coherent matter wave.  The answer to this question sets the motivation for our work here.  We will find that a field-theoretic description of matter waves provides an additional set of tools in which to treat atomtronic circuits. The approach leverages the analytical, heuristic, and numerical tools that have been well-developed for electromagnetics.  In particular is the appearance of ``Maxwell matter waves'' - the classical limit of coherent matter waves governed by matter-wave duals to Maxwell equations. Like their electromagnetic counterparts, Maxwell matter waves exhibit temporal coherence and they have several other aspects that are familiar from electromagnetics.

Here, we consider an oscillating current of neutral atoms, such as $^{87}$Rb, which repel each other due to mutual Van der Waals interactions. In field theory, atom interactions are accounted for by the introduction of a gauge field.  In comparison to the Coulomb forces between electrons, Van der Waals forces are weak and they are short-ranged. One might conclude that therefore the gauge field can be dismissed if, say, particles are far apart. But that conclusion proves to be misleading if not wrong.   The gauge field that embodies the interaction between neutral atoms is associated with energy and, generally, the transmission of power. \textcolor{black}{At low temperatures atoms interact through s-wave collisions, and thus the gauge field that characterizes these interactions assumes a relatively simple form. We will sometimes refer to this field as the ``matteron field'' to provide it  with an identity.  The matteron field, then, accounts for the interactions among the cold neutral atoms}

\textcolor{black}{With a motivation that parallels the electromagnetic case, we are particularly interested in the notion of a single-mode of the matteron field.  Generally a finite temperature gas, or even a pure BEC, will be comprised of a continuum of modes. }

Our theoretical development is carried out in three sections. The first is a classical, field-theoretic derivation of the family of Maxwell equations characterised by a few \textcolor{black}{formally-introduced} constants where the classical theory does not constrain the dynamical parameters of wave propagation. While assuming the constants to be the speed of light and free-space impedance leads to the well-known Maxwell equations, the same constants with different meaningful values can represent matter-wave duals to Maxwell's equations. These values are provided by a quantum mechanical treatment of the fields, which is carried out in the section that follows. This section also establishes the formal connection between the quantum-mechanical coherent state and the classical coherence of matter waves implied by Maxwell equations duals. With the dynamical parameters in hand, the third section returns to the classical treatment, giving the fully constrained Maxwell equation duals from which a quantitative correspondence between experiment and theory can be established. We close with a final Remarks section to highlight the similarities and differences between coherent matter-wave and electromagnetic wave theories.

\section{The Classical Matter-wave Field}

Our development of classical, non-relativistic field theory follows closely the excellent notes of Wheeler \cite{Wheeler.1995} which include substantially more discussion concerning the development than we provide. Typically,  modern physics treatments of field theory are written to address the domain of high-energy physics, so the constraint of Lorentz covariance and its tie to the speed of light is imposed early on \cite{Bailin.1993,Peskin.1995,Susskind.2017,Schwichtenberg.2020}. In the realm of ultracold atomic physics, particle velocities of 10 m/s are already very high, placing us in a non-relativistic realm. 

As a starting point, let us consider a Lagrangian density \cite{Wheeler.1995} characterising a flux of particles having mass $m$:
\begin{equation}\label{Eq:Lagragian0}
 \mathcal {L}_{0}(S,\partial S, R)=R \left\{ \left[ \frac{1}{2m} \boldsymbol{\nabla} S \cdot \boldsymbol{\nabla} S \right ] - U(\mathbf{x}) + (\partial_t S) \right \},
\end{equation}
in which $S$ is Hamilton's principal function, $U(\mathbf{x})$ is an applied particle potential, $R$ has units of $(\text{length})^{-3}$ and so describes the energy carried by the flux as a density, and $\partial_{i}$ is a partial derivative over time $t$ or spatial coordinates $i \in \{x,y,z\}$. \textcolor{black}{(Our symbol choice $R$ for density follows the notation of Wheeler\cite{Wheeler.1995}; the perhaps more natural $\rho$ is reserved for a later analog with electromagnetism.)} 

\textcolor{black}{
As an aside, we note that the Hamilton-Jacobi equation derived from the Lagrangian for $S$ is:
\begin{equation}
 \frac{1}{2m}\boldsymbol{\nabla} S \cdot \boldsymbol{\nabla} S - U(\mathbf{x})+(\partial_t S)=0.
\end{equation}
Considering for the moment a single particle, its momentum $\mathbf{p}= -\boldsymbol{\nabla} S$. So one can recognize in the above its relationship to the single-particle Hamiltonian:
\begin{equation}
 H(\mathbf{p},\mathbf{x})=\frac{1}{2m}\mathbf{p} \cdot \mathbf{p} + U(\mathbf{x}).
\end{equation}
}
The Lagrangian Eq. (\ref{Eq:Lagragian0}) is invariant under the global gauge transformation:
\begin{equation}
  S\rightarrow S^{\prime} = S + \ell w
\end{equation}
where the constant $\ell$ has dimensions of \textcolor{black}{action} and $w$ is dimensionless.

Establishing local gauge invariance begins with the introduction of vector and scalar gauge fields $\mathbf{A}$ and $ \phi$, and incorporates the replacements:
\begin{equation}
\begin{split}
 &\partial_{i} S \rightarrow \mathcal D S = \partial_{i}S - q A_{i}, \hspace{.5 cm} i\in \{x,y,z\}\\
 &\partial_t S \rightarrow \partial_t S + q \phi.
\end{split}
\end{equation}
Contriving to have the resulting equations as well as their units look familiar, here we have introduced \textcolor{black}{in a formal way} a ``charge'' $q$. \textcolor{black}{In electromagnetics the charge indicates the strength of the particle interactions; here the choice to include the symbol is gratuitous, since the interaction strength per particle or per mass could equally well be incorporated into the units for the gauge field.} The Lagrangian density is re-written:
\begin{equation}\label{Eq:L1}
 \begin{split}
& \mathcal {L}_{0}(S,\partial S, R) \rightarrow \mathcal {L}_{0}(S,\mathcal D S, R) \\
 &= R \left[ \sum_{i=x,y,z} \frac{1}{2m}\left( \partial_iS - q A_i\right)^2 -U(\mathbf x) +\left(\partial_t S + q \phi\right) \right]\\
 &\equiv \mathcal {L}_{1}(S,\partial S, R,\mathbf A,\phi).
\end{split} 
\end{equation}
The new Lagrangian is invariant with respect to a local gauge transformation characterized by $W(\mathbf{x},t)$:
\begin{equation}
 \left. \begin{gathered}
 S\rightarrow S^{\prime} = S + \ell W(\mathbf x,t)\\ 
 \mathbf A \rightarrow \mathbf A^{\prime}=\mathbf A + \frac{\ell}{q}\boldsymbol{\nabla} W(\mathbf x,t)\\
 \phi\rightarrow \phi^{\prime}=\phi -\frac{\ell}{q} \partial_t W(\mathbf x,t)\\
 \end{gathered} \right \}
\end{equation}
The Lagrangian Eq. (\ref{Eq:L1}) accounts for the energy of the particle flux as well as the energy of the interaction between the particles and the auxiliary fields $\mathbf A$ and $\phi$. We need to also incorporate the energy associated with the auxiliary fields by themselves. To that end, we introduce the matter-wave duals $\mathbf F$ and $\mathbf G$ of the electric field $\mathbf E$ and magnetic field $\mathbf B$:
\begin{eqnarray}
 & \mathbf F \equiv -\partial_t \mathbf A - \boldsymbol{\nabla}\phi, \\
 &\mathbf{G} \equiv \boldsymbol{\nabla} \times \mathbf A.
\end{eqnarray}
That these fields are gauge-invariant, as is the case for their electromagnetic duals, is easy to verify. With these fields so defined we can write the combined Lagrangian density written in a manner contrived to be later familiar:
\begin{equation}\label{Eq:TotalLagrangian}
 \mathcal{L}_c= \mathcal{L}_1 + \frac{1}{2} \left( \frac{1}{\eta_0} \mathbf{G} \cdot \mathbf{G} -\xi_0\mathbf{F} \cdot \mathbf{F} \right).
\end{equation}
We will have more to say about the matter wave duals $\xi_0$ and $\eta_0$ to the electric permittivity and permeability in Section IV.

The Hamilton-Jacobi equation for the field characterizing the density is:
\begin{equation}\label{Eq:H-JDensity}
 \partial_{t}R + \boldsymbol{\nabla} \cdot \frac{1}{m} R \left(\boldsymbol{\nabla} S - q \mathbf{A} \right)=0.
\end{equation}
This equation for the density expresses a continuity relation for the Noetherean current \cite{Schwichtenberg.2020}. Define charge and current densities:
\begin{eqnarray}
& \rho \equiv q R ,\\
& \mathbf{J} \equiv \frac{q}{m} R\left(\boldsymbol{\nabla} S - q\mathbf{A} \right).
\end{eqnarray}
We then have a familiar continuity relation:
\begin{equation}\label{Eq:Continuity}
 \boldsymbol{\nabla}\cdot \mathbf J + \partial_{t}\rho = 0.
\end{equation}

Since the divergence of a curl is identically zero and given the definitions of the fields $\mathbf{F}$ and $\mathbf{G}$:
\begin{equation}\label{Eq:MagField}
 \boldsymbol{\nabla} \cdot \mathbf{G}=0,
\end{equation}
\begin{equation}\label{Eq:FaradyLaw}
 \boldsymbol{\nabla}\times \mathbf{F}+\partial_{t}\mathbf{G} = 0.
\end{equation}
The Hamilton-Jacobi expression for the scalar potential is:
\begin{equation}
 \left[ \partial_{t} \frac{\partial}{\partial \left( \partial_{t} \phi\right)}+\boldsymbol{\nabla}\cdot \frac{\partial}{\partial \left(\boldsymbol{\nabla}\phi \right)}-\frac{\partial}{\partial \phi}\right] \mathcal{L}_c=0,
\end{equation}
which leads to:
\begin{equation}\label{Eq:GaussLaw}
 \boldsymbol{\nabla} \cdot \mathbf{F}=\rho / \xi_0 .
 \end{equation}
We have more to say about this Gauss' law dual in the Remarks section. Continuing, each component of the vector potential has a corresponding Hamilton-Jacobi equation, such as:
\begin{equation}
\begin{split}
 &\xi_0 \partial_t\left(\partial_t A_x+\partial_x \phi \right) + \frac{1}{\eta_0} \partial_x \left( G_y - G_z \right)\\
 & + \frac{q}{m} R\left( \partial_x S - q A_x\right)=0.
\end{split}
\end{equation}
Combining the spatial components and following the definitions of the fields and current leads to a final Maxwell equation:
\begin{equation}\label{Eq:AmpereLaw}
 \boldsymbol{\nabla}\times \mathbf{G}-\frac{1}{v_0^2}\partial_{t}\mathbf{F} = \eta_0 \mathbf{J}.
\end{equation}

We have now a complete set of Maxwell's equations: Eq. (\ref{Eq:GaussLaw}) as Gauss' law for the electric field, Eq. (\ref{Eq:MagField}) as Gauss' law for the magnetic field, Eq. (\ref{Eq:FaradyLaw}) as Faraday's law of induction, and Eq. (\ref{Eq:AmpereLaw}) as Ampère's circuit law with the addition of Maxwell's displacement current, along with the continuity equation Eq. (\ref{Eq:Continuity}). In Wheeler's words, in fact we have a family of Maxwell's equations \cite{Wheeler.1995}. Substituting $v_{0} \rightarrow c$, $\eta_0 \rightarrow \mu_0$, $\xi_0 \rightarrow \epsilon_0$ and of course $\mathbf{F} \rightarrow \mathbf{E}$ and $\mathbf{G} \rightarrow \mathbf{B}$ returns us to electromagnetics in familiar nomenclature. The charge and current densities are sources for the electric and magnetic fields. Equivalently we can view them as the source for the scalar and vector potentials.

Let us the introduce the matter-wave dual of the Lorenz gauge:
\begin{equation}
 \boldsymbol{\nabla} \cdot \mathbf{A}= - \frac{1}{v_{0}^2}\partial_{t} \phi.
\end{equation}
In this case a pair of wave equations govern the potentials:
\begin{eqnarray}
\left(- \boldsymbol{\nabla}^2 + \frac{1}{v_{0}^2}\partial_{t}^{2}\right) \phi=\rho / \xi_0, \\
\left(- \boldsymbol{\nabla}^2 + \frac{1}{v_0^2}\partial_{t}^{2}\right)\mathbf{A} = \eta_0 \mathbf{J}.
\end{eqnarray}
The velocity $v_0$ substitutes for the speed of light $c$, which is an experimentally determined fundamental physical quantity. The matter-wave dual will become evident as we blindly follow the course as we would for electromagnetics. In particular we are now in a position to consider an oscillating matter current. Consider \textcolor{black}{the positive frequency component of} a one-dimensional \textcolor{black}{oscillating} current:
\begin{equation}\label{Eq:CurrentWave}
 \mathbf{J}=\hat{x} J_{0} e^{i\left(k x - \nu t \right)},
\end{equation}
in which case the continuity equation, Eq. (\ref{Eq:Continuity}), insists:
\begin{equation}\label{Eq:DensityWave}
 \rho=\rho_0 e^{i\left(k x - \nu t \right)},
\end{equation}
along with:
\begin{equation}\label{Eq:ParticleDensity}
 \rho_{0}=\frac{k}{\nu} J_{0} = J_{0}\frac{1}{v_{\rm{m}}} ,
\end{equation}
where we define the velocity:
\begin{equation}
 v_{\rm{m}}\equiv \nu/k.
\end{equation}
\textcolor{black}{While $v_{\rm{m}}$ is the phase velocity associated with the current density, thinking forward to the quantum aspects, we can know in advance that it is also the group velocity of the particles \cite{Anderson.2021}.}   Let us correspondingly write:
\begin{eqnarray}
 \phi=\phi_{0} e^{i\left(k x - \nu t \right)},\\
\mathbf{A}=\hat{x} A_{0} e^{i\left(k x - \nu t \right)},
\end{eqnarray}
and further define a refractive index $n$:
\begin{equation}
 n\equiv \frac{v_0}{v_{\rm{m}}}.
\end{equation}
Then the amplitudes for the potentials are easily calculated:
\begin{eqnarray}
&&A_0 = \frac{\eta_0}{ k^2} \frac{n^2}{n^2-1} J_0, \\
&&\phi_0 = n v_0 A_0.
\end{eqnarray}
Borrowing once again from electromagnetics we write a matter-wave dual to Ohm's Law:
\begin{equation}\label{Eq:phi}
\phi_0 = Z_f J_{0}\frac{1}{k_0^2},
\end{equation}
where the basic wave number $k_0 \equiv k/n$ and the impedance is:
\begin{equation}\label{Eq:Impedance}
 Z_f\equiv -\sqrt{\frac{\eta_0}{\xi_0}} \frac{n}{1-n^2}.
\end{equation}

The Theory of Relativity regards the speed of light as an upper speed limit. We see through the impedance that there is also a speed limit: as $v_{\rm{m}}$ approaches $v_0$ so that the refractive index approaches unity, the field amplitudes diverge. The minus sign associated with the impedance has been explicitly brought out because, as we shall see in the next section, the matter wave speed limit is a lower bound rather than an upper bound, and the refractive index is typically less than unity rather than greater than unity as is the usual case for electromagnetics. That the impedance is negative has important physical significance, namely the power carried by the field:
\begin{equation}
 P_f \propto Z_f\left| J_0 \right|^2 ,
\end{equation}
is negative if the impedance is negative. Generally the concept of negative real-valued impedance is familiar in electronics, and there are a few interpretations that are equivalent. For the present, we recognize simply that the total power carried by the matter-plus-field is lower than the power carried by the matter \textcolor{black}{were there no interactions}. \textcolor{black}{The relationship between power, particle flux, and current is further discussed in Sec. \ref{Sec:coherence}.}

\begin{figure}[t]
\includegraphics[width=\columnwidth]{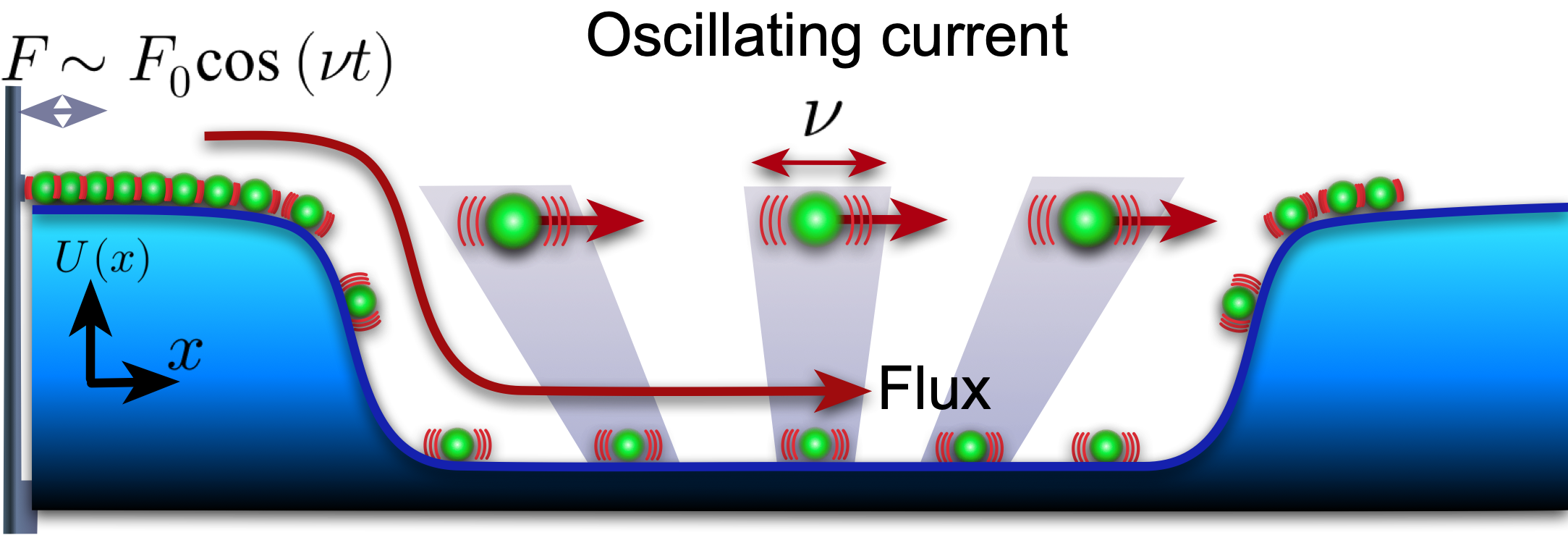}
\caption{\label{fig:Interactions} An illustration of the classical physics underlying gauge fields. Interactions among otherwise neutral atoms through van der Waals forces can be significant when atoms are close together, as in the left side of the figure. As atoms roll down the potential landscape their speed increases and so does the inter-atom distance, so that the interaction forces become very small; nevertheless, they must ``remember'' the consequences of the earlier interactions. An oscillating current imposed on the left will be preserved even as the atom flux moves from left to right so that it is coherent across the system. }
\end{figure}

To provide some insight into the formalism, let us consider a classical picture of particle dynamics as illustrated in FIG. \ref{fig:Interactions}. Plotted along the x-axis is the particle potential energy due to an applied potential. In the case of alkali atoms, sculpting a particular potential is conceptually simple to implement, say, utilizing laser light tuned to the red or blue side of an atomic resonance to either raise or lower, respectively, the atomic potential energy. An ensemble of atoms is initially on the left, biased with high potential energy. They are shoved from the left toward the right say, by the left-hand wall oscillating at the frequency $\nu$. The atoms fall down the potential, propagate along a relatively flat region, then climb back up the potential on the right. Of course they have gained speed as they roll down the potential, and therefore they are far apart, then move closer together again as they climb the potential.

Van der Waals forces are at play only when atoms are close together, therefore at the bottom of the potential their interactions are much weaker than at the top, where they are close together. Yet both the applied and the interaction potential are conservative, so the atoms will ``remember'' their interaction energy such that they return to the same configuration on the right as they began on the left (more or less, since we have not given detail on how the walls work). If one imagines a continual stream of atoms from the left, the average flux will everywhere be the same, as is the current oscillating everywhere along the system at frequency $\nu$. It is the gauge field that serves as the memory that invokes coherence of the current across the system.

\section{Quantum Treatment of the Matter-wave Field}

The field theory we have applied treats particles in terms of a delocalized field. This field, ``knows'' about the potential everywhere (say) in the $+x$ direction, and, although the field is everywhere in the half-space, it is also propagating in the $+x$ direction. In particular, in some sense its current position and velocity are dependent on a distribution of the possible past positions at historical times. The view of particles as delocalized entities suits well a transition to a quantum mechanical description. The notion of a short-range interaction is meaningful in a picture of localized particles undergoing a collision, yet when the particles are delocalized, in fact occupy an entire half-space, their interaction is manifest differently: if a steady oscillation is occurring in one location one should expect it to occur in all locations, loosely in keeping with the depiction in FIG. \ref{fig:Interactions}.

As a practical matter, absent from the matter-wave duals of Maxwell's equations are values of the two key parameters from which all dynamics follow. In electromagnetics the vacuum speed of light and impedance (or the vacuum permeability and permittivity) are taken as fundamental constants of nature and are determined empirically (as is the charge of the electron). To determine the dynamical constants of our field theory we must turn to a quantum-mechanical description, in which a Hamiltonian steps in for the Lagrangian and operators step in for dynamical coordinates and their corresponding functions and functionals.

The classical theory anticipates that the potential energy associated with the gauge fields is negative. Negative energy associated with quantization of the matter-wave gauge field is in stark contrast with the electromagnetic case. While the physics itself is straightforward, conceptual navigation through negative energy territory is best done with some care. We therefore begin with the basics.

The transition from the classical to the quantum description of matter is typically accomplished with the introduction of massive particle creation and annihilation operators $\hat{b}_m^\dagger$ and $\hat{b}_m$, together referred to as ``ladder'' operators \cite{Scully.1997}. They have the commutation property:
\begin{equation}
 [\hat{b}_m,\hat{b}_m^{\dagger}]=1 .
\end{equation}
In contrast to the classical picture illustrated in FIG. \ref{fig:Interactions}, the creation operator $b_m^\dagger$ creates a delocalized mode that exists everywhere in a half-space while it propagates in the $+x$ direction. \textcolor{black}{For the purposes of this Section, it can suffice to suppose that a mode propagates with a well-defined wavenumber, meaning that the particle potential is uniform. The operator treatment, however, equally well accommodates a spatially non-uniform potential, such as that depicted in  FIG. \ref{fig:Interactions}. } The Hamiltonian corresponding to a single-mode of the matter field is given by the number operator:
\begin{equation}\label{Eq:matterHamiltonian}
 \hat{H}_{m}=\hbar \omega_m \hat{b}_m^{\dagger} \hat{b}_m,
\end{equation}
for which the Planck-Einstein frequency is:
\begin{equation}
 \omega_m=\frac{\hbar k_{m}^2}{2 m},
\end{equation}
where $k_m$ is the deBroglie wavenumber associated with the particle momentum and corresponding particle group velocity:
\begin{equation}\label{GroupVelocity}
    v_{m} = \sqrt{2 \hbar \omega_m / m}.
\end{equation}
Eigenstates of the Hamiltonian are Fock (number) state: In particular the ground state energy is zero:
\begin{equation}
 \hat{H}_{m}\left| 0 \right>=0.
\end{equation}
Higher-lying Fock states are produced by repeated application of the creation operator:
\begin{equation}
 \left|N_m\right> = \frac{1}{\sqrt{\left(N_m+1\right)!}}(\hat{b}_m^\dagger)^{N_m}\left| 0 \right>,
\end{equation}
while the energy associated with a single mode is simply given by the number of particle excitations $N_{m}$ of the matter field:
\begin{equation}
 \hat{H}_{m}\left| N_{m} \right>=\hbar \omega_m N_{m} \left| N_{m} \right>.
\end{equation}

\textcolor{black}{The Hamiltonian Eq. (\ref{Eq:matterHamiltonian}) embodies the energy of a matter field, that is of the particles, alone.  It does not account for the particle interactions.  The classical development likewise began with the Lagrangian for the particles alone, then treated interactions through the introduction of a gauge field, and finally considered the energy of the gauge field alone.  } The quantization of the gauge field, i.e. the matteron field, is similar to that of the matter field. We introduce the ladder operators $\hat{a}_f^\dagger$ and $\hat{a}_f$, which obey commutations relations the same as those above for the matter, while they commute with each other, $[\hat{a}_f,\hat{b}_m ] = [ \hat{a}_f^\dagger, \hat{b}_m ] = 0$, etc.

The Hamiltonian associated with the matteron field, like the electromagnetic field, is itself analogous to that of the quantum-mechanical harmonic oscillator associated with harmonic frequency $\nu_f$, (for the time being, we take the oscillator frequency to be general, $\nu \rightarrow \nu_{f}$). A notable difference between the matter field and the matteron field is that the ground state of the latter has nonzero energy. More importantly, though, \textcolor{black}{is that self-consistency in the theory insists that the energy associated with the gauge field is negative, as commented upon in the Remarks section. } Glauber has pointed out that an inverted harmonic oscillator can be described by ladder operators identical to the normal oscillator, but is characterized by negative rather than positive energies \cite{GLAUBER.1986}:
\begin{equation} \label{Eq:MatteronHamiltonian}
    \hat{H}_{f} = -\hbar \nu_f\left( \hat{a}_f^\dagger \hat{a}_f  +\frac{1}{2} \right).
\end{equation}
The quantum of excitation of the matteron field is the dual of the photon of electromagnetics: a massless gauge boson.  We have referred to the dual elsewhere as a ``matteron'' \cite{Anderson.2021}, hence the naming of this gauge field.  In contrast to the photon, the matteron is associated with negative energy. 

As is the case for the matter field, the energy eigenstates of the matteron field are Fock states.  For a field comprised of exactly $N_f$ matterons:
\begin{equation}
        \hat{H}_{f}\left| N_{f} \right>=-\hbar \nu_f \left(N_{f}+\frac{1}{2}\right)  \left| N_{f} \right>.
\end{equation}

\textcolor{black}{As we move to explicitly consider particle interactions there is a subtle but important departure from the case of electrons interacting through the electromagnetic field.  In the latter case we can, philosophically at least, separate the photons from the electrons and then consider how they interact.  We saw from the classical treatment that an oscillating particle current serves as a source for the gauge field -one cannot exist without the other. For the corresponding quantum case, an excitation of a matter field is necessarily accompanied by an excitation of the matteron field.  } We are thus interested in the combined matter and matteron field excitations - in particular, what is appropriate to call ``a single-mode excitation of the matter-wave field.'' \textcolor{black}{Here forward we shall assume the excitation is one-for-one, i.e.} $N_{m}=N_{f}$.  In contrast to the familiar deBroglie matter-wave case, there are two distinct frequencies involved: the Planck-Einstein frequency $\omega_m$ associated with the particle and the frequency $\nu_f$ associated with the oscillating field, which is set by some external agent (a circuit, say) applying a time-varying force and causing a propagating oscillating current.  

We are now in a position to self-consistently determine that the limiting velocity $v_0$ is a lower bound on the particle velocity:  First, we can understand that the dynamics of the system must be such that the total energy is non-negative, meaning the particle energy must be greater than or equal to the matteron energy\textcolor{black}{, since a negative energy would imply that the system  lies in a bound state.}  This in turn means that the minimum velocity is dependent on the frequency $\nu_f$ associated with the gauge field:
\begin{equation}
    v_0 (\nu_f) \equiv v_{0f} = \sqrt{2 \hbar \nu_f / m}.
\end{equation}
Second, we see that physically the minimum velocity is such that the particle is never moving in the negative direction as it oscillates.  Doing so would contradict the assumptions, Eqns. (\ref{Eq:CurrentWave}) and (\ref{Eq:DensityWave}), that the charge density and current density are described by plane waves propagating in the positive $x$ direction.  \textcolor{black}{(On the other hand, particles moving in the negative $x$ direction is associated with plane waves of current and charge density propagating in the negative $x$ direction)} This is the rationale for clarifying the distinction between particle flux and particle current in the \ref{Sec:coherence}. Having a classical picture of an oscillating particle in mind, its motion, and hence particle flux, is always in the positive direction as long as its center of mass is moving sufficiently fast in the positive direction.  
Evidently:
\begin{equation}\label{Eq:RefractiveRatio}
    n = \sqrt{\frac{\nu_f}{\omega_m}}\equiv n_{\omega \nu}.
\end{equation}
In contrast to electromagnetics (and optics in particular), our refractive index is less than unity, $n_{\omega\nu}<1$, and often much less. We note that at the minimum particle velocity, i.e. $n=1$,  the wavelength $\lambda = 2 \pi/k = 2 \pi /(n k_0)$ associated with the field becomes equal to the particle de Broglie wavelength $\lambda =n 2 \pi/ k_0$.  

Given the multiple contexts of the noun ``field'' we clarify that we will utilize the operators $\hat{b}$ with the subscript $m$ or another subscript to refer specifically to the matter field, and the operator $\hat{a}_f$ specifically with the subscript $f$ to refer to refer to the matteron (gauge) field. The ``matter-wave field'' will refer to the two taken together. From here on the quantum-mechanical treatment we will focus on to the single-mode matter-wave case.  That means we shall fix the two frequencies, $\nu_f \rightarrow \nu$ and $\omega_m \rightarrow \omega$ and drop subscripts when there is no ambiguity.  Formally, matter and gauge fields occupy distinct Hilbert spaces.  In the single-mode case, however, the two are so tightly entangled that for the purposes of this work we can treat them together. We thus use $\hat{a}_\nu, \hat{a}_{\nu}^\dagger$ as matter-wave field ladder operators:
\begin{equation}\label{Eq:MWHamiltonian}
    \hat{H}_\nu =\hbar \left(\omega - \nu\right)\hat{a}_{\nu}^\dagger \hat{a}_\nu -\hbar\nu / 2.
\end{equation}
In particular for a Fock state \textcolor{black}{$N_f \equiv N_{\nu}$, }
\begin{equation}
    \hat{H}_\nu\left|N_\nu \right> = \hbar \left(\omega - \nu\right)N_\nu -\hbar\nu / 2.
\end{equation}

\begin{figure}[t]
\includegraphics[width=\columnwidth]{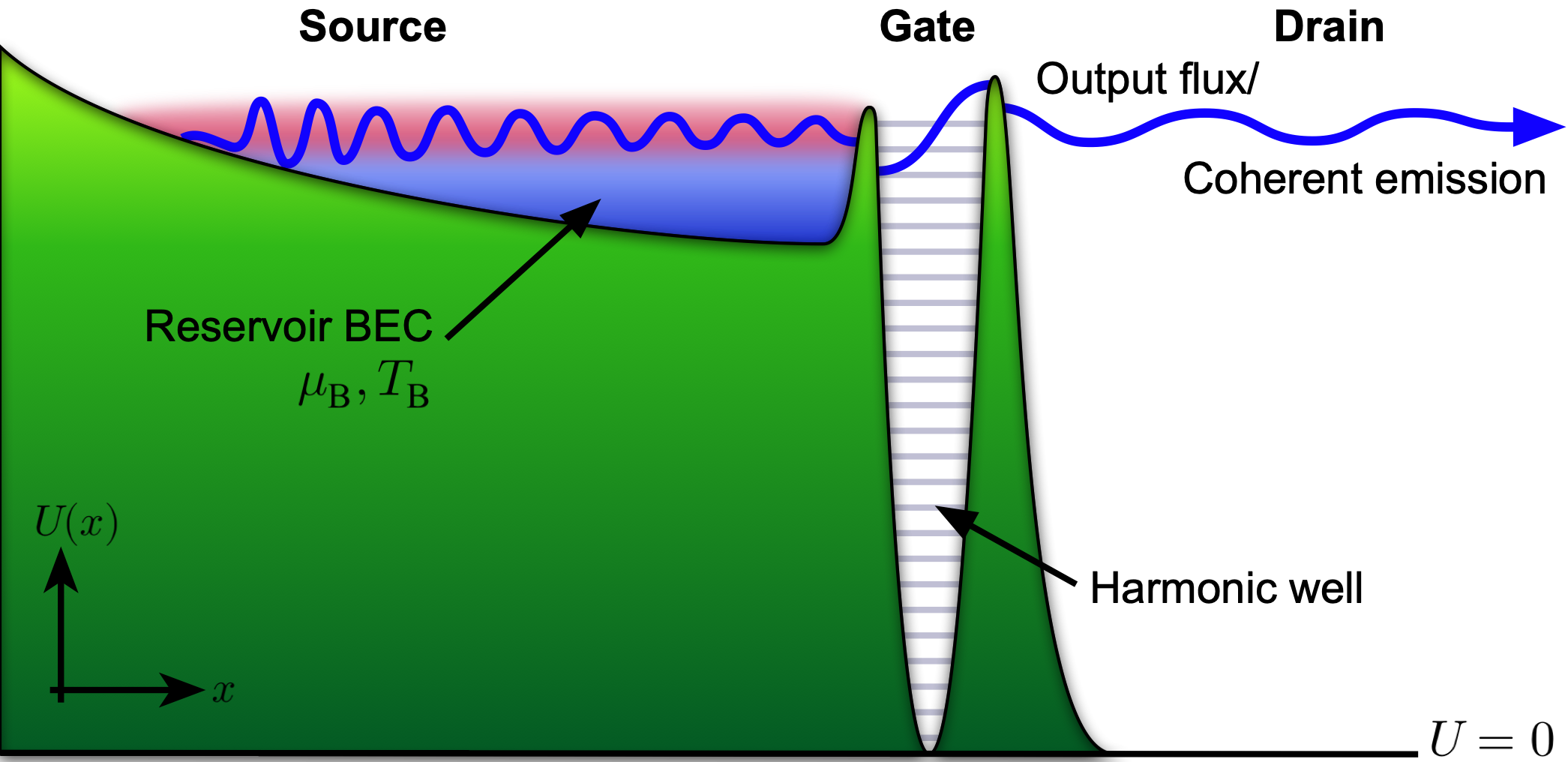}
\caption{\label{fig:Oscillator} The potential energy landscape for a coherent matter wave emitter. The source well contains a BEC that serves as a reservoir supply of particles.  For the purposes of this work, the gate well can be treated as a tuned element that selects the frequency $\nu$ of the oscillating current. \textcolor{black}{This configuration of potential energy and atoms is a self-oscillating system\cite{Anderson.2021}.  The matter-wave emission is due to system dynamics rather than an imposed oscillating force as notionally depicted in Fig. {\ref{fig:Interactions}}}. Reproduced with permission from D. Z. Anderson, Physical Review A 104, 033311 (2021). Copyright (2021) under a Creative Commons License.  }
\end{figure}

Next we seek to formalize a means of exciting the matter-wave field from some ground state.  In order to allay concerns of creating mass from vacuum, \textcolor{black}{for simplicity} we a reservoir that has no matteron \textcolor{black}{energy ($\nu_f=0$)}, but provides a supply of massive particles  having energy $\hbar \omega$. We are thinking in particular of a Bose-condensate having a large and fixed number $N_r$ of identical particles  as such a supply (see FIG. \ref{fig:Oscillator}).   we set as the system single-mode ground state:
\begin{equation}
    \left|0_{r\nu} \right> \equiv \left|N_r \right> \left| 0_{\nu} \right>.
\end{equation}
As a condensate, we take all particles to having identical energy $E=\hbar \omega$.  Let us refer to the two sub-spaces comprising our system in terms of reservoir states ($r$) and matter-wave states ($\nu$), as in practice the former will involve trapped particles and the latter particles (massive and matterons) that propagate in free space.  We introduce joint operators:
\begin{eqnarray}
    \hat{c}_{r\nu}\equiv \hat{b}_r^\dagger \hat{a}_{\nu},\\
    \hat{c}_{r \nu}^\dagger \equiv  \hat{a}_{\nu}^\dagger \hat{b}_r,
\end{eqnarray}
such that:
\begin{equation}
    \hat{c}_{r \nu} \left|0_{\omega \nu}\right> = 0,
\end{equation}
and, for example,
\begin{equation}
    \hat{c}_{r \nu}^\dagger \left|0_{r \nu}\right> = \sqrt{N_r} \left| N_r-1 \right> \left| 1_\nu \right>.
\end{equation}
such that the reservoir gives up a particle to the matter-wave field.  Higher excitation of the field is accomplished with repeated action of $\hat{c}_{r\nu}^\dagger$.  The joint operators entangle the reservoir and propagating fields. In the limit that the reservoir particle number is large it can be treated as a constant, and we can drop the reservoir terms to recover the matter-wave Hamiltonian, Eq. (\ref{Eq:MWHamiltonian}). 

Total system energy is, or course, not conserved, since for every particle contributed by the reservoir, $\hbar \nu$ of energy is removed by the gauge field.  We keep in mind that the emitted massive particles themselves each carry the full $\hbar \omega$ worth of energy of the particle contributed by the reservoir -it is the negative energy carried by the matteron that is responsible for the energy shortage.  This indicates that the emission \textcolor{black}{of the matteron field itself} induces cooling of the reservoir\textcolor{black}{, although the emission of the particles proves to cause heating \cite{Zozulya.2013} such that the next effect is heat added to the reservoir } The detailed physics of these  process we have omitted.  

The familiar classically coherent wave corresponds to the quantum mechanically pure coherent state given as the eigenstate of the annihilation operator. 
\begin{equation}\label{Eq:CoherentState}
    \hat{a}_\nu \left|\alpha_\nu \right> = \alpha_\nu \left| \alpha_\nu \right>,
\end{equation}
in which the eigenvalue is complex, $\alpha_\nu = \left|\alpha_\nu \right| e^{i \varphi_\nu}$, and the phase is associated with the oscillation frequency $\nu$.  The expectation of the energy carried by the coherent state is:
\begin{equation}
\begin{split}
     \left<E_{\nu}\right> &= \hbar \left(\omega - \nu\right) \left|\alpha_\nu\right|^2 - \hbar\nu/2 \\
     &=  \hbar \nu  \frac{1-n^2}{n^2}  \left|\alpha_\nu\right|^2 - \hbar\nu/2.
\end{split}
\end{equation}

Our theoretical development has involved no reference to the origin of the particle interaction - only that it exists and can be characterized in terms of particle mass $m$ and charge $q$.  The pivotal requirement is that the interaction energy is sufficient to supply a matteron having energy $\hbar \nu$.  In FIG \ref{fig:Oscillator}, this corresponds to a BEC having an interaction mean-field contribution to its chemical potential $\mu_B \geq \hbar\nu$ in order to supply matterons to the emission. 

Finally, noting the role of impedance in the previous section, we introduce the quantum impedance:
\begin{equation}
    Z_0 \equiv  \hbar /q^2.
\end{equation}
In closing we note that labeling of the charge $q$ is convenient for the sake of the analogy and for dimensional bookkeeping.  In practical applications, however, one will find it convenient to set the charge either to the particle mass or to unity, corresponding to the impedance given in units of (energy-per-mass)-per-(mass-per-second) or in units of (energy-per-particle)-per-(particle-per-second). 

\section{Coherent Matter-Wave Parameters}\label{Sec:coherence}
\begin{figure}[t]
\includegraphics[width=\columnwidth]{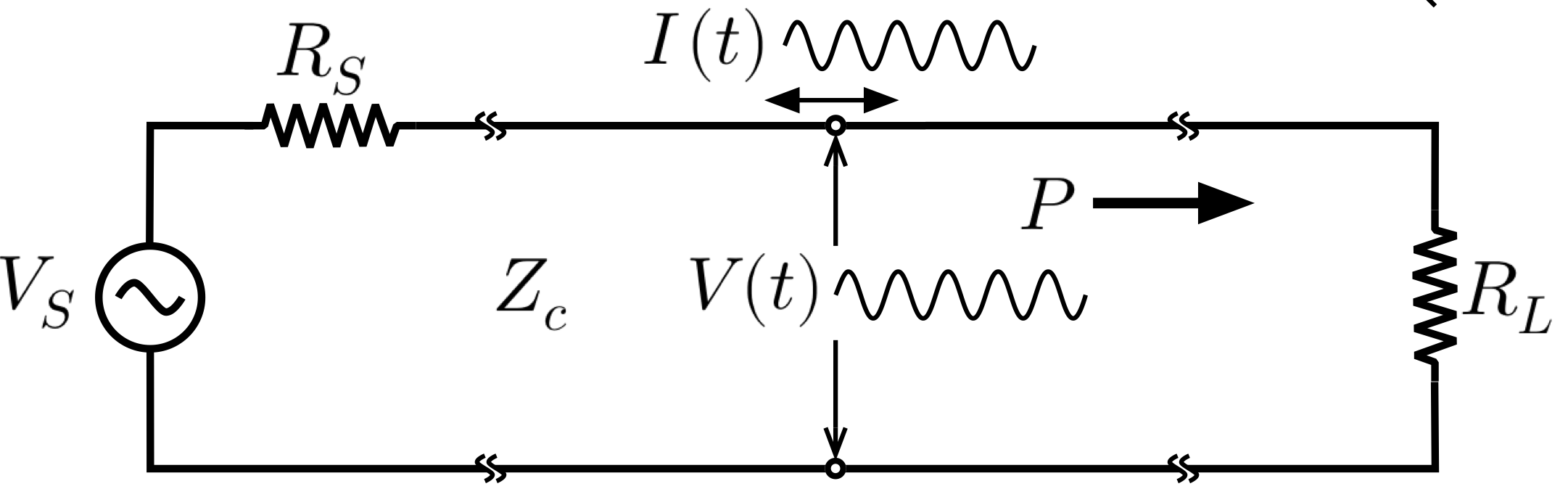}
\caption{\label{fig:TransmissionLine} A microwave transmission line circuit is useful for highlighting the distinction between current and flux. The flux corresponds to the flow of power in the circuit. The maximum power $P$ available from a microwave source $V_S$ is delivered to a load when the source resistance, load resistance, and characteristic impedance of the transmission line are all equal, $R_S = R_L=Z_c$. In such a case the power and the atom flux flowing from the source to the load is constant while the voltage across and current through a given point along the line vary sinusoidally.  In particular, the direction of the current through a point flows alternately towards the load and towards the source. }
\end{figure}

\textcolor{black}{The analogy with electromagnetics can provide insight into the matter-wave system through a simple radio-frequency circuit. FIG. \ref{fig:TransmissionLine} shows an oscillating microwave voltage source delivering power to a load $R_L$ through an electromagnetic transmission line \cite{Gonzalez.1997}. The voltage and current waves carry energy, and if the source, load resistance, and transmission line impedance are equal, all the energy is delivered to the load. Of note is the fact that the electric current oscillates between negative and positive values and thus flows in both directions.  The voltage oscillates in phase with the current;  as a consequence, the power flows continuously from left to right.}

\textcolor{black}{The example highlights an important distinction between current and flux: We will use the term ``flux'' to refer to the flow of power, equivalently in the case of atomtronics the flow of particles, or quantum-mechanically to the flow of probability. As with the transmission line example, we consider a case in which ultracold atom flux is continuous, while atomtronic current and voltage are oscillating amplitudes. By contrast, a BEC whose center-of-mass is oscillating within a trap is described by an oscillating flux.} 

Impedance reflects the energy cost of adding an additional charge to the current. The quantum impedance determines the duals to permittivity and permeability, since, in analogy with electromagnetics:
\begin{eqnarray}
    Z_f =- \sqrt{\eta/\xi } ,\\
    v_{\rm{m}} =\frac{v_0}{n} = 1/\sqrt{\eta \xi }.
\end{eqnarray}
This leads us to:
\begin{equation}\label{Eq:permeability}
\begin{split}
     & \eta_0 \equiv  Z_0  \sqrt{\frac{ m}{2 \hbar \nu}}=  Z_0 \frac{1}{v_0 }, \\
     & \eta = \eta_0 \frac{n^2}{1-n^2} .
\end{split}
\end{equation}
\begin{equation}\label{Eq:Permittivity} 
\begin{split}
     &\xi_0 \equiv  \frac{1}{Z_0}\sqrt{\frac{ m}{2 \hbar \nu}}=\frac{1}{Z_0} \frac{1}{v_0}, \\
     &\xi =  \xi_0 \left(1-n^2\right).
 \end{split}
\end{equation}
Finally:
\begin{equation}
    Z_f=-\frac{\hbar}{q^2}\frac{n}{1-n^2}.
\end{equation}

Having the impedance and velocity in hand allows one to calculate all field amplitudes given the current density amplitude $J_0$.  In particular the potential is given by Eq. (\ref{Eq:phi}) while the remaining field amplitudes are:
\begin{equation}
\begin{split}
    A_0 &= - \frac{1}{n\nu}\left|Z_f \right| J_0/k_0,\\
    F_0 &= -i Z_0 J_0/k_0, \\
    G_0 &= -i \frac{1}{\nu}\left| Z_f \right|  J_0.
\end{split}
\end{equation}

That the gauge and matter fields propagate together leads one naturally to think in high-frequency electronic circuit terms, in which currents and fields also coexist.  In this context it is convenient to work using the duals of electric voltage and current.  To that end we introduce:
\begin{eqnarray} 
    \mathcal{V}_0 =k_0^2 A \phi_0 \\
    \mathcal{I}_0= A J_0,
\end{eqnarray}
where A is an effective area over which the fields and currents are transversely confined.  The relationship between the two is set by $Z_f$, which in this context is referred to as the characteristic impedance \cite{Morse.1953,Gonzalez.1997} such that,
\begin{equation}
    \mathcal{V}_0 = Z_f \mathcal{I}_0.
\end{equation}

While the current $\mathcal{I}_0$ is an amplitude, generally it is the flux $I_{\rm{m}}$, defined as the number of massive particles per second that traverse a given position, which is straightforward to measure. We are interested particularly in the energy stored in and carried by the gauge field. The power carried by the gauge field is,  simply:
\begin{equation}
    P_f = -I_{\rm{m}} \hbar \nu = \frac{1}{2} Z_f \left|\mathcal{I}_0\right|^2.
\end{equation}
Hence in the limit $n \ll 1 $
\begin{equation}\label{Eq:Current}
    \mathcal{I}_0 = \frac{q}{\sqrt{n}}  \sqrt{2 I_m  \nu} ,
\end{equation}
\begin{equation}\label{Eq:Potential}
    \mathcal{V}_0 =- \hbar\frac{\sqrt{n}}{q} \sqrt{2 I_m  \nu} .
\end{equation}

In circuit applications it will often be convenient to consider the energy of the particles $\hbar \omega$ carrying the currents as fixed, while the oscillation frequency $\nu$ is taken to be variable.  In such cases it will be convenient to make the substitution $k_0 \rightarrow k/n$ in the various formulas for fields and impedance.

\section{Remarks}
Through field theory we have derived a set of matter-wave duals to Maxwell's equations describing electromagnetic waves. We have intentionally assembled the theory to resemble electromagnetics; a benefit, because of the valuable intuition, heuristics, and analytical tools that can be transferred from one domain to the other, but also a detriment, as it obscures the contrasts between the two. 

Conspicuously absent in the treatment are the duals to the constitutive relations corresponding to $\mathbf{D}=\epsilon \mathbf{E}$ and $\mathbf{H} = \mathbf{B}/\mu$.  In particular for example, the gauge field energy in the Lagrangian Eq. (\ref{Eq:TotalLagrangian}) uses the duals to the vacuum permeability and permittivity.  That is because here there are no duals to polarization or magnetization of a medium.  The velocity associated with the gauge field is tied to that of the atoms, which in turn is influenced by an external potential that acts directly on the atoms.  By contrast, with electromagnetism, the gauge fields $\mathbf{E}$ and $\mathbf{B}$ themselves interact with a medium in which they propagate. 

We have shown that an oscillating matter current gives rise to a coherent matter wave that obeys the Maxwell equations duals. Of dramatic difference from the electromagnetic counterpart is that the oscillating matter current does not produce a radiated field that exists in a source-free region.  On the contrary, \textcolor{black}{through the negative sign in the Hamiltonian Eq. (\ref{Eq:MatteronHamiltonian}) for the matteron field} we see that the matter and its gauge field are bound together.  The quantization of the field introduces the matteron, a massless gauge boson that is dual to the photon. In a particle picture the matteron is bound to the massive particle, while the classical particle can be pictured as undergoing longitudinal oscillation. \textcolor{black}{ It is worth noting that if the matteron energy were taken to be positive, it would not be bound.  As a massless boson, the matteron would then necessarily travel at the speed of light (see for example Schwichtenberg\cite{Schwichtenberg.2020}), and be indicative of a long-range force, not of the short-ranged Van der Waals forces at play in our particle interactions  }

It is reasonable to consider whether our matter waves are in fact sound waves, and whether the matteron should in fact be called a phonon.  Sound waves, such as those that can propagate in a BEC \cite{Pethick.2002}, arise from the interaction among the condensate atoms and the propagation velocity depends upon, among other things, the atomic density.  We have seen in the discussion at the end of the previous section that the interaction energy associated with a matter wave as it propagates in otherwise empty space can be arbitrarily low, and that the wave propagation speed is given by the particle group velocity. So there is nothing acoustic about the matter waves under consideration.  

The picture of longitudinal oscillation of the massive particle is perhaps at first objectionable because oscillation implies a restoring force.  While matterons are massless they carry momentum $\hbar k_0$ and they are bound to the massive particle with energy $\hbar \nu$.  Thus we can think of a massive particle attached to a lighter one with a spring that supplies a restoring force as the two oscillate with respect to each other.  

The contrast between matter waves and electromagnetics is further highlighted, for example, by the dual to Gauss' law, Eq. (\ref{Eq:GaussLaw}) which might suggest field lines emanating from an isolated charge as it does for the electric case. Yet the dispersion implicit in the dual to permittivity $\xi$ Eq. (\ref{Eq:Permittivity}) belies the complexity under the hood, so to speak, particularly at DC.  In short, at a distance away from any current no matteron field is detectable. Yet within the flux of particles their effects are measurable: that the energy of the bound particle-plus-matteron state is less than the energy of the massive particle alone reveals the physics behind the flow of negative power carried by the gauge field.  

The dynamical parameters for matter waves, namely the propagation speed and impedance, are imposed by quantum mechanics rather than by Relativity as is the case for electromagnetism.  The quantum treatment of the fields provides a clear understanding of the meaning of ``classically coherent matter wave'' as an eigenstate of the combined matter-field-plus-gauge-field annihilation operator, which has a well-defined phase relative to the oscillation frequency $\nu$ of the current. Keeping to the classical domain, it seems appropriate to refer to these waves as ``Maxwell matter waves''.  In this context, our duals to Maxwell's equations were contrived to appear the same as the familiar set, yet the fundamental conclusions are that the propagation velocity has a lower rather than upper limit, and that the impedance of free space has a negative real value for Maxwell matter waves, compared with the positive 377 ohms for electromagnetic waves.

At the same time, Maxwell matter waves do obey their dual Maxwell's equations, and thus the heuristic and analytical tools that already exist for electromagnetics can be put to effective use to address matter-wave calculations.  Indeed one can work with the fields $\mathbf{F}$ and $\mathbf{G}$, but it is also natural to work in a circuit theory context involving scalar potential and current, connected to each other through impedance, as we have seen.  Step changes in the refractive index, for example, lead to familiar coefficients of wave amplitude and power reflection and transmission that arise in the case of particle wavefunctions, of light waves, of microwaves, and so on  \cite{Morse.1953,Hecht.2017,Gonzalez.1997}.  

 
It is relevant to note that the interference of Maxwell matter waves reveal substantially different character than the interference of de Broglie matter waves.  For example, the fringe spacing of interfering de Broglie waves will decrease with increasing particle velocity while that of Maxwell matter waves will increase.   

Traditional is the de Broglie view in which thinks of matter classically as a particle phenomenon whose wave character is revealed by quantum mechanics.   Here we see that Maxwell matter waves provide the complementary view in which matter can be viewed classically as a wave phenomenon whose particle aspects are revealed by quantum mechanics.

\begin{acknowledgments}
The work of K. Krzyzanowska  was supported by the Quantum Science Center (QSC), a National Quantum Information Science Research Center of the U.S. Department of Energy (DOE). The work of D.Z. Anderson was supported by Infleqtion. We are grateful to S.Du and V. Colussi for valuable discussions. 
\end{acknowledgments}

\section*{author declarations}
\subsection*{Conflicts of Interest}
Dana Z. Anderson has stock in, and serves on the Board of Directors of ColdQuanta Inc., dba Infleqtion.

\subsection*{Data Availability Statement}
Data sharing is not applicable to this article as no new data were created or analyzed in this study.

\bibliography{MMW}

\end{document}